\documentclass[a4paper, russian]{article} 
\usepackage[T2A]{fontenc} 
\usepackage[cp1251]{inputenc}
\usepackage{graphicx}

\begin{document}

\title{THE FUNCTIONAL INTEGRAL IN THE HUBBARD MODEL } 
\author{V.M. Zharkov \\ 
\\ Institute for Natural Sciences,  \\ Perm State University, Perm, Russia, e-mail: vita@psu.ru.}
\maketitle

\begin{abstract} In a new functional integral approach proposed for the model, we find the regime with a deformed in­tegration measure in which the standard integral is replaced with the Jackson integral. We indicate the relation to a p-adic functional integral. For the magnetic and electronic subsystems in the effective functional that results from the operator formulation of the Hubbard model, we find the two-parametric quantum derivative resulting in the appearance of the quantum $SU_{rq}(2)$ group. We establish the relation to the one-parametric quantum derivative and to the standard derivative. 
\end{abstract}

\section{Introduction}

\bigskip 

The Hubbard model \cite{Hubbard}   remains the main object on which to probe new approaches to strongly cor­related systems. It is traditionally assumed that effects of strong Coulomb repulsion play an important role in understanding mechanisms of high-temperature superconductivity, the physics of a metal-dielectric transition of the Mott-Hubbard type, and the related magnetic states \cite{Izyumov}. The experimental discovery of spin liquid in an organic crystal  $(BEDT-TTF)_{2}Cu_{2}(CN)_{3}$ \cite{BEDTTTF} 
made the situation even more complicated, establishing the necessity of taking spin liquid into account in the phase diagram of the model under study. It recently became clear that we need new approaches for help, in particular, in understanding properties manifested by materials and separate molecules used in nanotechnology devices. Materials with a one- or two-dimensional conductivity band, for instance, polymers, are widely used in these devices. For the TA-PPE polymer (the abbreviation of poly(p-phenylene ethynylene), jumps of two types, step and sawtooth, were observed \cite{c14}  in the dependence of conductivity on the applied voltage.

\begin{figure}[ht]
\centering
\includegraphics*[width=300pt, height=200pt]{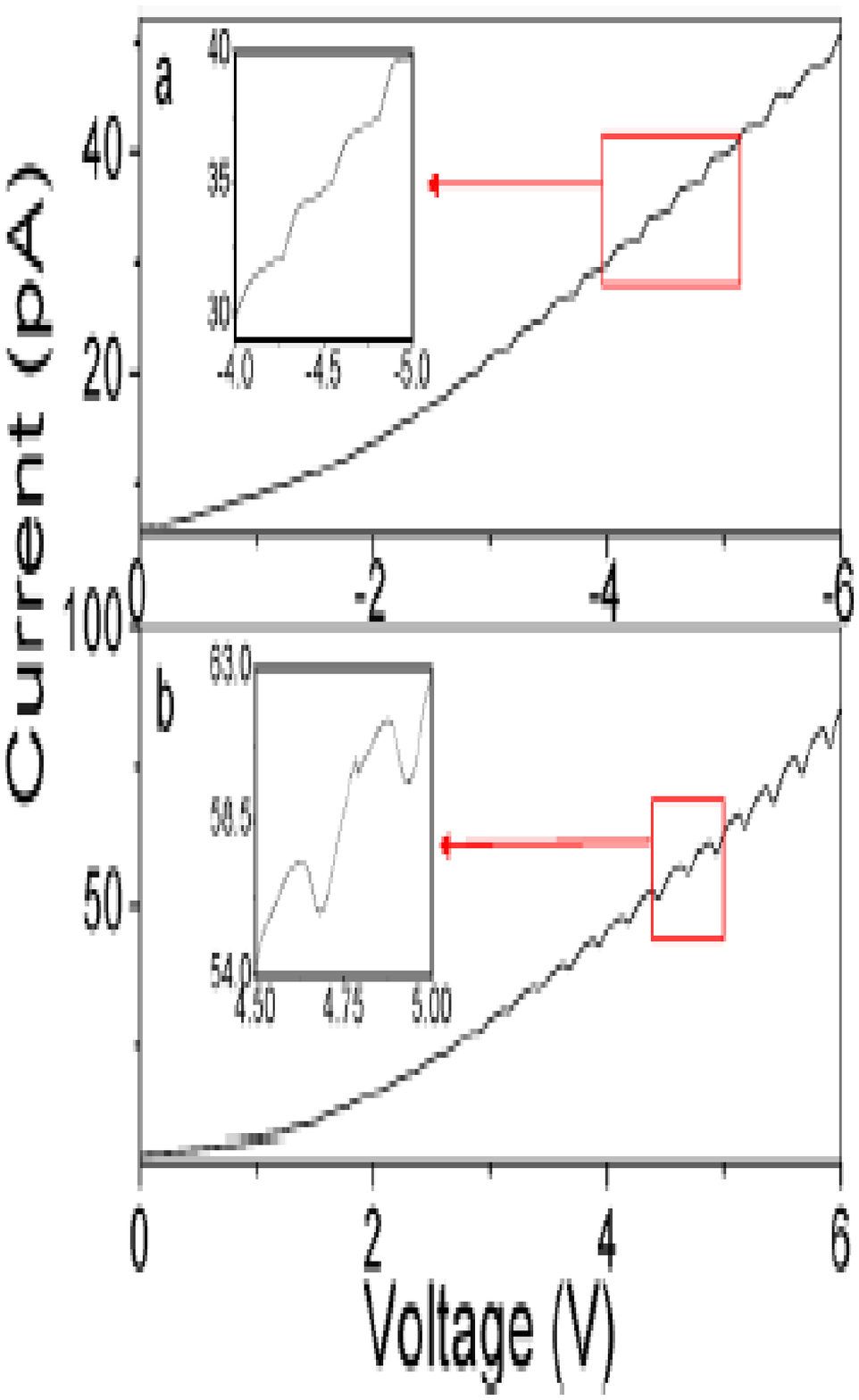}
\caption{Fig. 3.  The dependence of the conductivity on the voltage for the polymer TA - PPE: in the insets, we present magnified jumps in the ranges (a) from -4.0 V to -5.0 V and (b) from 4.5 V to 5.0 V.
}\label{fig:name4}
\end{figure}

Jumps of the conductivity and magnetization were observed in other systems and materials \cite{c11,c12,c13}.  We proposed \cite{cc1,cc2}  describing such a behavior of macroscopic quantities using functions of the argument taking values in the p-adic number field. The argument of this function determines the value of the voltage, while the numerical simulation has demonstrated that the function itself, which depends on a one parameter, results in the two experimentally observed dependence types: step and sawtooth. Increasing the accuracy when calculating this function of the p-adic argument (we took the values $ p = 2,3$ for the p-adic numbers) revealed the appearance of a nested systems of jumps in the conductivity behavior. The physical pattern of such a behavior in strongly correlated systems was described in \cite{cc2} on the qualitative level. It was mentioned that in the regime of strong electron repulsion in one-dimensional systems, the electron system splits into fluctuating clusters containing several electrons each. As the temperature and voltage increase, the viscous motion of these clusters taken as a whole results in processes of splitting them into smaller clusters. We thus obtain a hierarchical embedding of jumps of smaller amplitude into the structure of jumps with larger amplitude.

This paper is devoted to justifying that a regime in which the studied macroscopic quantities are de­scribed by functions of a p-adic argument can arise in strongly correlated multielectron systems with strong Coulomb repulsion. We here justify the following scenario: p-adic numbers appear through a deformation of the integration measure depending on the deformation parameter q. The Jackson measure used when calculating macroscopic quantities appears in the functional integral in the strong correlation regime. In the case $q = 1/p$, where p is a prime, the Jackson integral becomes the p-adic integral \cite{Vladimirov}. We can therefore use functions of a p-adic argument to describe jumps in the Hubbard model.
In particular, we here continue the investigation of the new formulation of the functional integral that we proposed in \cite{Tower,ZharkovKirchanov}. We note that this approach differs substantially from functional formulations of the Heisenberg and Hubbard models common in the foreign literature. Therefore, in what follows, we present a comparison with different versions of functional representations for creation-annihilation operators proposed in various papers. We show that the obtained expressions for supercoherent states result in more complex composite expressions for the operator symbols than those provided by the slave-boson and slave-roton approaches \cite{Hermele,Kim1}. The expressions for the creation-annihilation operators obtained here with various group reduction schemes in the Hubbard model first proposed in \cite{Tower}, result in all currently known formulations of the functional integral for systems with strong Coulomb interaction. Moreover, as shown in \cite{ZharkovKirchanov}, the proposed formulation allows studying various cohomologies of groups and supergroups and thus provides a controlled concretization and expansion of dynamical groups and symmetries spontaneously appearing when strengthening the interaction in multielectron systems.
Our proposed representation for the effective functional and creation-annihilation operators is essen­tially nonlinear. This nonlinearity allows segregating those terms that provide quantum derivatives deter­mining generators of the quantum algebra in expressions belonging to the universal enveloping algebra. In \cite{Zharkov1}, we introduced the approximation in which radius vectors of vector fields were independent of the dynamical field coordinates. This assumption is equivalent to the approximation in which the values of the fields that correspond to the Casimir operators and to the invariants of the classical groups $SU(2)$ and $ SU(1,1)$ in our formulation are coordinate independent.
Below, we demonstrate that these combinations determine two parameters of deformation of the quan­tum Lorentz group, which is the group of four-dimensional rotations. We treat the transition to the one-parameter quantum derivative in detail and show that the magnetic and electric subsystems of the Hubbard model are described by deformed versions of nonlinear sigma models. We further calculate the part of the functional integral for the Hubbard model that results from expressions containing this quantum derivative.
We realize our consideration on model contributions naturally present in the Hubbard model. We consider the limiting cases with respect to the deformation parameter and calculate the functional integral in these limit regimes.
We calculate the effective functional following from the kinetic energy of the Hubbard model using supercoherent states containing no more than the first powers of fermionic fields. We fix the dynamical fields such that the residual expressions indicate the difference between our approach and other approaches in the literature. The main distinct feature of our representation is its nonlinearity, which allows introducing a two-parameter quantum derivative into the problem. When passing to a one-parameter quantum derivative, we obtain a deformed nonlinear sigma model in the auxiliary dimension determined by the scale factor. Further, by analogy with the diagram technique for atomic X-operators \cite{Izyumov, Tower}, we consider the integral of the scale factor linear in the quantum number. This integral determines the integration measure. Directly calculating the corresponding series, we show that in the limit case as the deformation parameter tends to zero, this series converges to the Jackson integral.
Obtaining deformations of the integration measure when varying the deformation parameter is our main result. In the limit case of small $q$, the standard integral becomes the Jackson integral.

\section{The effective functional of the Hubbard model}

We briefly describe the general scheme for constructing the effective functional for the Hubbard model proposed in \cite{Zharkov1} and recently further developed, for example, in \cite{ZharkovKirchanov}. We begin with the Hubbard model written in terms of the standard creation-annihilation operators,

\bigskip

\begin{equation} H=-W\sum_{ij\sigma }\alpha _{\sigma ,i}^{+}\alpha _{\sigma ,j}+U\sum_{i,\sigma \
}n_{\sigma ,i}n_{-\sigma ,i}-\mu \sum_{\sigma ,i}n_{\sigma ,i}, \label{h1} \end{equation}

where $\alpha_{\sigma ,i}^{+},\alpha_{\sigma ,j}$  are the creation and annihilation operators,  $n_{\sigma
,i}$ is the electron density operator, $W,U,\mu $  are the respective width of the conductivity band, the one-site repulsion of two electrons, and the chemical potential, and $ \sigma $  determines the spin value. We sum over the indices $i$ and $j$ labeling the lattice sites. For instance, for a one-dimensional lattice, we have  $—N/2 < i < N/2$, where $N$ is the total number of atoms. For electrons, we sum over the indices $ \sigma=\pm 1/2 $. In the weak correlation regime with a small on-site repulsion, we take the zeroth approximation for the kinetic energy and treat the repulsion perturbatively. In the "atomic" approach with strong repulsion, we take a one-site contribution of the form
$$U n_{\sigma ,i}n_{-\sigma ,i}-\mu n_{\sigma ,i}.$$

as the zeroth approximation.
In what follows, we omit the index $i$ in order to concentrate on the internal structure of the appearing fiber bundle. This term is diagonal in the "atomic" basis and has the eigenfunctions and eigenvalues $\epsilon$:
$$|0\succ ;|+\succ =\alpha_{\uparrow }^{+}|0\succ; |-\succ =\alpha_{\downarrow }^{+}|0\succ;|2\succ
=\alpha_{\uparrow }^{+}\alpha_{\downarrow }^{+}|0\succ , $$
$$ \epsilon_0=0, \epsilon_{+}=-\mu, \epsilon_{-}=-\mu, \epsilon_2=U-2 \mu .$$

Calculating the matrix elements for $\alpha_{\sigma }^{+},\alpha _{\sigma }$  in this basis, we obtain the following matrix representation.
For instance,    $\alpha _{\uparrow }^{+}$ is:
\begin{equation}
 \alpha _{\uparrow }^{+}=\left(\begin{array}{cccc}
 0 & 0 & 0 & 0 \\
 1 & 0 & 0 & 0 \\
 0 & 0 & 0 & 0 \\
 0 & 0 & -1 & 0
\end{array}\right) =X^{+0}-X^{2-}.
\label{op}
\end{equation}
The operators $\alpha _{\sigma}^{+}$  therefore contain two nonzero matrix elements. We introduce the operators $ X^{rs},r,s=0,+,-,2 $ containing only one nonzero matrix element.   
We then obtain the expansion of the creation and annihilation operators in terms of the Hubbard operators \cite{Hubbard}:

$$\alpha _{\uparrow }^{+}=X^{+0}-X^{2-},\alpha _{\downarrow }^{+}=X^{-0}+X^{2+},\alpha _{\uparrow
}=X^{0+}-X^{-2},\alpha _{\downarrow }=X^{0-}+X^{+2}.$$

Because we have a four-dimensional basis and the creation and annihilation operators are expressed as 4x4 matrices, we have a basis comprising 16 operators in the general case. From this set we remove the unity operators and the operators of the form

$$\gamma _5=\left( \begin{array}{cccc}
  1 & 0 & 0 & 0 \\
  0 & -1 & 0 & 0 \\
  0 & 0 & -1 & 0 \\
  0 & 0 & 0 & 1
  \end{array} \right). $$

We use this notation for this operator because it coincides with the $ \gamma_5 $ operator written in the chiral basis in quantum field theory (QFT). It also equals

$$ \gamma_5=(X^{00}-X^{22})^2-(X^{++}-X^{--})^2. $$

We can express the one-site Hubbard repulsion in terms this operator. Because we can express it in terms of other operators, we do not take it into account in what follows. The remaining operators on a site in the given basis can then be separated into the fermionic operators of the form
$$(X^{0+},X^{0-},X^{+0},X^{-0},X^{+2},X^{-2},X^{2+},X^{2-}),$$

and the bosonic operators of the form

$$(X^{+-},X^{-+},X^{++}-X^{--},X^{02},X^{20},X^{00}-X^{22}).$$
We omit the lattice index of these operators to indicate the possibility of obtaining generators of the global dynamical algebra. We can endow the Hubbard operators with the lattice index by taking $N$  their copies, where $N$ is the total number of lattice sites, and constituting the direct product from these copies.
A contemporary review of the atomic approach was presented in \cite{Irkhin}.  In terms of these operators, the Hubbard model becomes
\begin{equation} H=U\sum_{i,r}X_{i}^{rr}-W\sum_{ij\alpha \beta }X_{i}^{-\alpha }X_{j}^{\beta }
\label{HubbAtomic} \end{equation}

In what follows, we use the functional formulation for multielectron systems proposed in \cite{Zharkov1}. The evolution operator between the initial and final states in this formulation is given by the following functional integral with the action expressed in terms of the effective functional, which is to be calculated using operator expression (\ref{HubbAtomic}) using the supercoherent states (we note that the index labeling the lattice in this formulation is lacking for the set of operators$ X^{rs} $, which therefore constitute the set of generators of the dynamical superalgebra):

\begin{equation} <G_{f}|e^{-iH(t_{f}-t_{i})}|G_{i}>=%
\int_{|G(t_{i})>=|G_{i}>}^{|G(t_{f})>=|G_{f}>}D(G,G^{\ast })e^{-iS[G,G^{\ast }]}; \label{fi}
\end{equation}

The action here has the form

\begin{equation}
 S[G,G^{\ast
}]=\int_{t_{i}}^{t_{f}}dt\int_{V}d^3r\frac{<G(r,t)|i\frac{\partial }{\partial
t}-H|G(r,t>}{<G(r,t|G(r,t>} \label{lag}
\end{equation}

and the integration measure is

$$D(G,G^{\ast }) = \prod_{t_{i}<t<t_{f}} \prod_{r} dG(r,t)^{\ast }dG(r,t)<G(r,t)|G(r,t)> $$.

The integrand in QFT formula (\ref{lag}) is called the Lagrangian. We avoid using this term wherever possible because in contrast to QFT, we begin with an already secondarily quantized representation and subsequently pass to an effective model containing composite and nonlinear fields. Such a passage is nontrivial, and to use the term "Lagrangian," it is desirable to know the Lagrangian equations of motion at least. Unfortunately, in our approach, for each representation of the chosen supercoherent state, we must separately investigate the search for the variables and their conjugate momenta. We therefore use the term "effective functional" (although it is not very precise) instead of the term "Lagrangian."

We use a function equal to the supercoherent state of the form

\begin{equation}
\mid G>=\exp \left( \begin{array}{cccc} E_{z} & 0 & 0 & E^{+} \\ \chi _{1} & h_{z} & h^{+} & 0 \\
\chi _{2} & h^{-} & -h_{z} & 0 \\ E^{-} & -\chi _{3} & \chi _{4} & -E_{z}\end{array}\right) \mid
0> \label{Scs}
\end{equation}
to describe local properties of the strongly correlated system. The exponent in this expression contains dynamical fields depending on the coordinates and time. We determine the electric field by the three-dimensional vector,

\bigskip $$\mathbf{\vec{E}}=(E^{+}(x,y,z,t),E^{-}(x,y,zt),E^{z}(x,y,z,t)).$$
magnetic fields has three components depending on the space-time ordinates,

$$\mathbf{\vec{h}}=(h^{+}(x,y,z,t),h^{-}(x,y,z,t),h^{z}(x,y,z,t)).$$

The fermionic fields are described by odd Grassmann-valued functions of the coordinates and time,

$$\chi_{k}^{\ast }(x,y,z,t),\chi_{k}(x,y,z,t),k=1,2,3,4.$$
Expression (\ref{Scs})  is the solution of the following problem: in the atomic basis, the site state is described by a wave function of the form
\begin{equation}
\mid G \left( i\right)> =a_{i}\mid i0\rangle +b_{i}\mid i+\rangle +c_{i}\mid i-\rangle +d_{i}\mid
i2\rangle , 
\label{func}
\end{equation}
 where $a_{i} ,b_{i},c_{i} ,d_{i} $ - are the coefficients of the atomic basis functions depending on the index  $i $.
 We have the normalizing condition for the wave function

$$ < G \mid G> ^{2}=a_i^{2}+b_i^{2}+c_i^{2}+d_i^{2}=1$$.

The drawbacks of using functions of this type are well known. A function of the type$a_{i} $  is symmetric with respect to its coordinates (bosonic type), while the functions $b_{i} ,c_{i}$  are skew-symmetric under coordinate interchanges (fermionic type). The general form of the functions $ \mid G > $ is expression (\ref{Scs}), where the exponent has the form $ \sum_{mn} X^{mn} \phi_{mn} $,and the functions $ \phi_{mn} $  are $ \chi_k^{\ast}, \vec{h},\vec{E}. $ Supercoherent state (\ref{Scs}) determines the supergroup representation with the set of supergenerators $ X^{nm}. $
We note that the expressions for the above supercoherent states were calculated exactly in \cite{Zhar2}. As noted in \cite{ZharkovKirchanov}, using computers and computer programs for working with supermatrices and superpoles, we can exactly calculate the functional itself. It is clear from \cite{Zhar2} that the main distinction between our approach and the approaches in other papers is the nonlinear form of the supercoherent state expression. Be­low, we show that all the representations proposed in the literature for symbols of the creation-annihilation operators follow from our representation when the fields are fixed. We also show that it is the essentially nonlinear nature of the expressions in the supercoherent state, which are functions of electric and magnetic fields, that distinguishes our representation from the previously proposed functional integral representa­tions for the Hubbard model. These nonlinear expressions contain quantum derivatives and, as a result, quantum groups. We note that after the super-Yangian structure in the one-dimensional Hubbard model was discovered \cite{Uglov}, there were numerous attempts to find quantum groups in the Hubbard model, but they were unsuccessful. We use an absolutely different approach in this paper, and the discovery of quantum structures in the Hubbard model is therefore explainable.
The expression for the effective functional $S$ governs the transition from the operator formulation in terms of the creation-annihilation operators and the Hubbard operators to the field theory formulation in terms of dynamical fields of the Bose and Fermi types. As shown in \cite{ZharkovKirchanov}, the main point of this transition is the passage to the problem formulation in terms of the superbundle determined by a representation of the local supergroup of four-dimensional space rotations. In what follows, we need only the symbols of the creation and annihilation operators determined by the expressions $<G\mid \alpha _{\sigma }^{+}\mid G>,<G\mid \alpha _{\sigma }\mid G>.$  The general expressions for supercoherent states were calculated in \cite{Zhar2}. Here, we take expressions for these states in the zeroth- and first-order approximations with respect to the Grassmann fields. We write the expressions for the coherent states and their conjugates in these approximations:

$$\mid G>=\left( \begin{array}{c} ch(E)+E_{z}\frac{sh(E)}{E} \\ a_{1}^{+}\chi _{1}+a_{2}^{+}\chi
_{2} \\ a_{1}^{-}\chi _{1}+a_{2}^{-}\chi _{2} \\ E^{-}\frac{sh(E)}{E}\end{array}\right) $$

\bigskip

$<G\mid =(ch(E)-E_{z}\frac{sh(E)}{E},(a_{1}^{+})^{\ast }\chi _{1}^{\ast }+(a_{2}^{+})^{\ast }\chi
_{2}^{\ast },(a_{1}^{-})^{\ast }\chi _{1}^{\ast }+(a_{2}^{-})^{\ast }\chi _{2}^{\ast
},-E^{+}\frac{sh(E)}{E})$

\bigskip

The action of the operators on this state leads to the expressions

\bigskip

$\alpha _{\uparrow }^{+}\mid G>=\left( \begin{array}{c} 0 \\ ch(E)+E_{z}\frac{sh(E)}{E} \\ 0 \\
-a_{1}^{-}\chi _{1}-a_{2}^{-}\chi _{2}\end{array}\right) ,\alpha _{\downarrow }^{+}\mid
G>=\left( \begin{array}{c} 0 \\ 0 \\ ch(E)+E_{z}\frac{sh(E)}{E} \\ a_{1}^{+}\chi _{1}+a_{2}^{+}\chi
_{2}\end{array}\right) ,\alpha _{\uparrow }\mid G>=\left( \begin{array}{c} a_{1}^{+}\chi
_{1}+a_{2}^{+}\chi _{2} \\ 0 \\ -E^{-}\frac{sh(E)}{E} \\ 0\end{array}\right) ,\alpha
_{\downarrow }\mid G>=\left( \begin{array}{c} a_{1}^{-}\chi _{1}+a_{2}^{-}\chi _{2} \\
E^{-}\frac{sh(E)}{E} \\ 0 \\ 0\end{array}\right). $

\bigskip

The symbols for the creation and annihilation operators calculated on these states and written in the spinor form are

$$ \left( \begin{array}{c} <G\mid \alpha _{\downarrow }^{+}\mid G> \\ <G\mid
\alpha _{\uparrow }^{+}\mid G>\end{array}\right) =E_{12}^{`}a_m\chi +\hat{E}_{11}A_m\chi ^{\ast },
$$

\begin{equation} \left( \begin{array}{c} <G\mid \alpha _{\uparrow }\mid G> \\ <G\mid \alpha
_{\downarrow }\mid G>\end{array}\right) =\hat{E}_{22}am \chi +E_{21}^{`}Am\chi ^{\ast }. \label{symbol}
\end{equation}

Here, the 2x2 matrices $\hat{E}$  and $\hat{h}_i $  are

\bigskip $E_{12}^{`}=\hat{E}_{12}\left( \begin{array}{cc} -1 & 0 \\ 0 & 1\end{array}\right)
,E_{21}^{`}=\hat{E}_{21}\left( \begin{array}{cc} -1 & 0 \\ 0 & 1\end{array}\right) ,$ $$\left(
\begin{array}{cc} \hat{E}_{11} & \hat{E}_{12} \\ \hat{E}_{21} & \hat{E}_{22}\end{array}\right)= \left( \begin{array}{cc}
ch(E)+E_z sh(e)/E & E^{+}sh(E)/E \\ E^{-} sh(E)/E & ch(E)-E_z sh(E)/E\end{array}\right)$$
$$a_m=f_3 (ln(f_3/f_2))' +f_2((ln(f_2))'+E_z)\hat{h}_2. $$ $$ \hat{h}_2= \left( \begin{array}{cc} (ln(f_2))'+h_z
& h^{+} \\ h^{-} & (ln(f_2))'-h_z\end{array}\right)$$

$$\sigma=\left( \begin{array}{cc} 0 & 1 \\ 1 & 0\end{array}\right); A_m=\sigma a_m^{*}; \chi=(\chi_1,\chi_2)^T; \chi^{*}=(\chi_1^{*},\chi^{*})^T . $$

We recall the fields over which we integrate in the functional integral: the vectors $\bf{\vec{E}}$ and $ \bf{\vec{h}}$ describing fluctuations of the respective electric and magnetic degrees of freedom and the Grassmann fields $\chi_i $.  The combinations  $ E=\sqrt{E_z^2+E^{+}E^{-}}, h=\sqrt{h_z^2+h^{+}h^{-}} $, which are invariants of the two $SU(2)$ groups, also enter the supercoherent state. We here set them to be coordinate- and time-independent constants. These constants determine two deformation parameters in the quantum group problem. We now consider the coefficients  $ f_2$ and $f_3$. Expressions for these quantities are nonlinear in the above invariants. The method for calculating them exactly was developed in \cite{Zhar2}. In the nonlinear representation of supercoherent states, we have four such coeficients:$ f, f_2,f_3,f_4$. We describe the trick used to calculate these coefficients in [17]. We multiply the invariants $ E$    and $ h$ in the formulas by the scale factor x and subsequent show that we can obtain $ f, f_2,f_3,f_4$   from $ f$  by taking derivatives with respect to $ x$. At the end of the calculations we set the parameter $x$ equal to unity. In the formulas presented above, such differentiations are marked by the prime.
Expression (\ref{symbol}) determines the representation for the creation-annihilation operators in the functional integral in terms of the three components of the vector $ \vec{h} $. The representation depending on the vector $ \vec{h} $ was proposed in \cite{Hermele}. We note that in the expressions there and also here, the local group is $SU(2) * SU(2)$.
We consider particular cases of the above composite expressions for the creation-annihilation operators:
1) If $(E^{+},E^{-},E_{z})=0,(h^{+},h^{-},h_{z})=0$ ,then: $<G\mid \alpha _{\sigma
}^{+}\mid G>=\chi _{\sigma }^{\ast },<G\mid \alpha _{\sigma }\mid G>=\chi_{\sigma} $ i.e., we obtain the holomorphic representation for the creation-annihilation operators. Substituting them in (\ref{h1}), we obtain the standard representation for ferminonic systems in terms of odd Grassmann-valued fields \cite{Popov}
\bigskip 2) If $(h^{+},h^{-},h_{z})=0,$ $(E^{+},E^{-},E_{z})\neq 0$ then expression (\ref{symbol}) generates the following formulas for the functional symbols of the creation and annihilation operators:

\begin{equation} \left( \begin{array}{c} <G\mid \alpha _{\downarrow }^{+}\mid G> \\ <G\mid \alpha
_{\uparrow }^{+}\mid G>\end{array}\right) =E_{12}^{`}f_3 B\chi +\hat{E}_{11}f_3 B^{*}\sigma\chi ^{\ast
}, \end{equation}

\begin{equation} \left( \begin{array}{c} <G\mid \alpha _{\uparrow }\mid G> \\ <G\mid \alpha
_{\downarrow }\mid G>\end{array}\right) =\hat{E}_{22}f_3 B \chi +E_{21}^{`}f_3 B^{*} \sigma\chi ^{\ast
}  \end{equation}

where $ B=(ln(f_3))' + E_z. $ Setting the expression for /$f_{3}$ equal to unity, we reduce these expressions to the slave-roton representation in \cite{Kim1}.

3) If $(h^{+},h^{-},h_{z})\neq 0,$ $(E^{+},E^{-},E_{z})= 0$ , then expression (8) generates the following formulas for the functional symbols of the creation and annihilation operators\cite{Kim1}:

\begin{equation} \left( \begin{array}{c} <G\mid \alpha _{\downarrow }^{+}\mid G> \\ <G\mid \alpha
_{\uparrow }^{+}\mid G>\end{array}\right) = \hat{E}_{11}A_m\chi ^{\ast }, \end{equation}

\begin{equation} \left( \begin{array}{c} <G\mid \alpha _{\uparrow }\mid G> \\ <G\mid \alpha
_{\downarrow }\mid G>\end{array}\right) =\hat{E}_{22}a_m \chi \end{equation}

$a_m=\hat{h}_3 $, and in the logarithmic function is $f_3$, in this formula, we have  $A_m=\sigma a_m^{*}. $  We have thus shown that each representation for the creation-annihilation operators in the Hubbard model known to us is contained in some particular case of expression ((\ref{symbol}) . We note that the main feature distinguishing our representation from those in the literature is the presence of the coefficients $f_{2}$ , $f_{3}$ и $f_4$. Expressions for these coefficients were given in \cite{Zhar2}. Below, we clarify the consequences of the presence of these coefficients.

\section{ Quantum derivatives in the Hubbard model}

We see which functional results from the operator expression for the magnetic subsystem of the Hubbard model. Substituting expressions for the creation-annihilation operators in the expression for the kinetic energy, we obtain the functional

$$ H=\sum \limits_{ij}(f_{4})^2\left( \begin{array}{c} \chi_{1}^{*} \\ \chi _{2}^{*}\end{array}%
\right)_{i}^T \left( \begin{array}{c} \chi_{1} \\ \chi _{2}\end{array}\right)_{j} $$

We note that for /$f_{4}=1$, our representation coincides with the slave-roton representation used in many papers. Our aim is to study those terms in the functionals that are absent in the cited papers. We take the unit matrix as the matrix field $ \hat{E}$, which indicates the independence on the space coordinates. We mentioned in the introduction that we want to obtain the expression for the coherent state for such $ \hat{E}$. Because $ \chi $  is a fermionic field, when passing to the momentum representation, we obtain a half-filled Fermi band over which we integrate to obtain the total number of fermions, i.e., a constant. We substitute this constant for the quadratic combination of fields $\chi _{i}^{\ast }\chi _{j}$.  As the result, we obtain the expression for the effective functional at coinciding arguments $(i=j) $ :
\begin{equation} H=(f_{4})^{2} \label{Kinetic} \end{equation}

 where the coefficient $ f_4$ is \cite{ZharkovKirchanov}: $ f_4=(E sh(E)-h sh(h))/(E^2-h^2). $
 This is precisely the factor that distinguishes our functional from the one in \cite{Hermele,Kim1}. At first glance, it is just a nonlinear coefficient. But...!
We introduce the two-parameter derivative acting on a function $ f(x)$ by the formula

$$D_{rq}f(x)=(f(rx)-f(qx))/((r-q)x)$$.

It is known \cite{Chakrabarti}, that the three operators $(D_{rq},x,x\frac{\partial }{\partial x})$ constitute the two-parameter quantum algebra \ $SU_{rq}(2)$  with the following commutation relations for the generators $(S^{+},S^{-},S_{z})$ = $(D_{rq},x,x\frac{\partial
}{\partial x})$ :

$$\lbrack S_{z},S^{+}]=S^{+},[S_{z},S^{-}]=-S^{-},S^{+}S^{-}-\frac{r%
}{q}S^{-}S^{+}=[2S_{z}]_{r,q}$$  We have the definition of the two-parameter quantum number \cite{Chakrabarti}:

$$[X]_{rq}=\frac{q^{X}-r^{-X}}{q-r^{-1}}$$

Using the expression for $D_{rq}$  we can rewrite the nonlinear coefficient $f_{4}$ in the form

$$ D_{E^{2},h^{2}}(x)(\sqrt{x}sh{\sqrt{x}}/x) =(E \sqrt{x}sh(\sqrt{x}E)-h
\sqrt{x}sh(\sqrt{x}h))/(x(E^{2}-h^{2})). $$

We introduced the separate parameter $x$ into the expression for $ f_4$. The representation of this quantity in \cite{Zhar2} implies that x determines the scale factor and is the dilaton field. We now show this. The complete set of fields in expression (\ref{Scs} for the supercoherent state and the functional integral integration measure is determined by the superalgebra whose matrix generators enter (\ref{Scs}. It does not contain the dilation generator, whose matrix representation is $\gamma_5$ in the chiral representation related to the third diagonal term in (\ref{Scs}. To take this term into account, we must calculate matrix (6) exactly with the full set of generators. Unfortunately, this problem is much harder than the problem solved in\cite{Zhar2}. We introduce the scaling factor for the local fields $\bf{\vec{E}}$ and $\bf{\vec{h}} $  such that $ \bf{\vec{E}}``->\beta \bf{\vec{E}}, \bf{\vec{h}}``->\beta \bf{\vec{h}}$  and pass to the primed fields. This factor can be a local $ x$-dependent field $ \beta(x)$, ) or a global $ x$-independent field. The field j3 is complementary to the fields $\bf{\vec{E}}$ and $\bf{\vec{h}} $ and must enter the nonlinear expressions for the supercoherent state. We can introduce a functional integration over this field, and this field then enters the integration measure. Because fields of this type (of the dilation field in QFT) had never been previously discussed in the Hubbard model framework, we below concentrate on discussing possible qualitative effects resulting from taking this field into account. In this paper, we set the scaling factor independently of the coordinate $x$. We have already shown that the quantum derivatives act on precisely this variable. We note that $ \beta(x)$,  is replaced with $x$ in formula (\ref{Kinetic}) in order to introduce a unified notation for the entity of the extended space-time coordinates. Taking the scaling factor into account means passing to the $(d+l)$-dimensional space-time in the case of the $d$-dimensional initial space-time. Substituting the derivative $D_{rq}$ in expression (\ref{Kinetic}), we obtain the functional of the form

\begin{equation} H=\sum \limits_{x}(D_{rq}(x)f(x))^{2}  \label{magnet} \end{equation}

where  $f(x)$ is given by expression in the previous formula on which the derivative acts. The difference from the nonlinear sigma model is due to specific properties of the coordinate $x$, which defines the dilaton field. We consider the limit as r$ r\rightarrow q$ or $E^{2}\rightarrow 1/h^{2}$. It is clear that $D_{rq}\rightarrow D_{q}$ in this limit and becomes the one-parameter quantum derivative acting on functions by the formula
$$D_{q}f(x)=(f(qx)-f(q^{-1}x))/((q-q^{-1})x).$$
The commutation relations for the operators ($(D_{q},x,x\frac{\partial }{\partial x})$ constitute the quantum group $SU_{q}(2)$ \cite{Pressley} with the commutation relations

$$[S_{z},S^{+}]=S^{+},[S_{z},S^{-}]=S^{-},[S^{+},S^{-}]=\frac{sh(S_{z}\ln (q))}{sh(\ln (q))}.$$
Letting $q\rightarrow 1$,  we reduce the action of the quantum derivative $D_q$ on arbitrary functions $f(x)$ to the action of the standard derivative \cite{Katz}, i.e., $D_{q}\rightarrow \frac{\partial }{\partial x}$ . Replacing $D_{q}\rightarrow \frac{\partial }{\partial x}$, we obtain the standard nonlinear sigma model $(\frac{\partial f(x)}{\partial x})^{2}$. We note that $ f$ is a solution of the normalizing constraint condition on the coherent state, which customarily defines a vector field of unit length. Such a model describes a magnetic subsystem in a functional integral approach. On the operator level, it is the Heisenberg magnet in the Hubbard model. This magnet model is defined on the dilaton field and is one-dimensional. Returning to the general effective functional in the Hubbard model, we note that we have found a quantum symmetry and the way to introduce it into strongly correlated systems. It is especially important that subsequent contractions of quantum derivatives coincide with three levels of the tower of symmetries discovered in the Hubbard model in \cite{Tower}. Formula (\ref{magnet}), implies that in multielectron systems, we must study "quantum" magnets whose variables are generators of quantum Hopf algebras, not simple Lie algebras.
We very briefly indicate the differences in the physical behaviors of such magnets. We here discuss only the behavior of the magnetic momenta and discuss more serious differences later. It is known that the
Casimir operator $S_{z}^{2}+S^{+}S^{-}$ of the group $SU(2)$ take the value: $S(S+1)$. The Casimir operator of the group $SU_{q}(2)$ is
$S^{+}S^{-}+(\frac{sh(qS_{z})}{q})^{2}$ and  it is close to zero as  $q->1$. In the physical language, this means that at the phase transition point at which a nonzero value of the parameter q appears, spin can arise smoothly with its value changing continuously from 0 to 1/2. This picture of the Mott-Hubbard-type metal-dielectric transition seems more natural compared with the phase transition scenario in which the spin instantly jumps from 0 (metal) to 1/2 (insulator of the Mott-Hubbard type). We now pass to studying a qualitative  a behavior of a functional of type (\ref{magnet}) using the functional integral.

\section{ The Jackson integral}
The above analysis of nonlinear expressions in the supercoherent state demonstrates that we can write complicated expressions in terms the action of the quantum derivative containing one or two deformation parameters. These derivatives act on functions of the variable, which was introduced in \cite{Zhar2} as a purely technical tool ensuring a uniform form of writing expressions for $ f_2, f_3, f_4$. The same variable was again used purely technically in \cite{ZharkovKirchanov} to calculate and write nonlinear expressions.
In the preceding section, we showed that by taking the scaling factor into account, we obtain a model resembling a nonlinear sigma model but with quantum derivatives instead of the standard ones. In the limit where the one-parameter quantum derivative transforms into the standard derivative, we obtain the standard nonlinear sigma model. We thus qualitatively obtained a fact that is well known in the Hubbard model: the magnetic properties of the Mott-Hubbard dielectric at large values of the Hubbard repulsion $U$ are described by the Heisenberg model with the exchange integral equal to $ W^2/U $ \cite{Izyumov}. 
Below, we use the diagram technique for spin and Hubbard operators and also our diagram technique for constructing the "tower of symmetries" [\cite{Tower}. When constructing the diagram technique for the Heisenberg model, we found the following trick useful. To the model action, which is quadratic in the spin operators, we add a term that is linear in the spin operators and proportional to $ S_z$, the so-called Zeeman contribution. We then study the quadratic terms perturbatively. Summing the diagrams responsible for the local ordering results in a nonzero contribution to the magnetic field and ensures the possibility of calculating means over the ground state when setting the previously introduced field to zero. Our scheme and expression(\ref{magnet}), imply that we encounter exactly the same situation. We cannot take (\ref{magnet}), as the leading approximation in contrast to the case where we have the standard, not quantum, derivative. We must add a linear derivative and study how it affects calculating the ground state means. Below, we show that by deforming the derivative, we do not introduce new interaction types, as could be expected, but rather deform the functional integration measure, i.e., we change the type of the functional variable itself. In addition to the standard bosonic fields described by complex-valued functions and fermionic fields described by odd Grassmann-valued functions variables, fields with a nontraditional integration measure, for instance, taking values in the field of p-adic numbers, appear in the functional integral. The base of this field is determined by the means of the bosonic electric and magnetic fields, which provide the deformation parameters for the quantum groups.
The standard scheme for studying functional integral (\ref{fi}) begins with the "zeroth" approximation ob­tained from an effective functional quadratic in either bosonic or fermionic variables, which reduces to Gaussian integrals over complex variables parameterized by momenta. We must be able to evaluate func-tionals containing quantum derivatives. For subsequent use, we briefly describe the scheme for constructing an effective functional using a holomorphic representation for creation and annihilation operators. To define an integral over trajectories completely, we must construct a functional series, which can then become a formal perturbation series. For this, we must know how to calculate integrals of the form
\bigskip \begin{equation} \int_0^{c} f(x)e^{-bx^{2}}dx=\sum \limits_{i=0}^{\infty
}\frac{f^{(n)}(0)}{n!}\int_0^{c} x^{n}e^{-bx^{2}}dx \label{series} \end{equation}
for an arbitrary function $f(x)$.
Clearly, we can exactly evaluate Gaussian integrals with power-law dependences, which allows stan­dardly constructing a formal perturbation series. Using this, for each bosonic quantum system with a Lagrangian quadratic in the fields, we can set the complex numbers $\varphi ,\varphi ^{\ast }$ into correspondence with the creation-annihilation operators $\beta _{\sigma }^{+},\beta _{\sigma },$ satisfying the standard commutation relations. Each functional integral with a bosonic Lagrangian containing interactions of fields can then be constructed as a formal se­ries in the interaction constant. To study effective functionals of type (\ref{magnet}), we must replace the exponential in expression (\ref{series}) as

$$\exp (b\varphi \varphi ^{\ast })\rightarrow \exp (bxD_{q})$$

Because we replace the pair of operators $(x,\frac{\partial }{\partial x})$ with the holomorphic coordinates $(\varphi ,\varphi ^{\ast })$, we obtain our variant of quantization if we substitute the quantum derivative  $D_{q}(x)$. for the standard one. Prom the definition of  $D_{q}$, we have

$$D_{q}\rightarrow \frac{sh(\ln (q)x)}{sh(\ln (q))x}=\frac{[x]_{q}}{x}$$

Substituting this expression in the exponent results in the expression
\begin{equation} \int_0^{c} f(x)e^{b[x]_{q}}dx  \label{Qseries} \end{equation}
where \ $[x]_{q}=\frac{q^{x}-q^{-x}}{q-q^{-1}}=\frac{sh(x\ln (q))}{sh(\ln (q))}.$
 Calculating a functional integral contain­ing general quantum derivatives $D_{rq}$, in a simpler case of the quantum derivative $D_{q}$, therefore requires calculating integrals of form (\ref{Qseries}), with arbitrary functions $f(x)$ of the variable $x$. We have thus described the scheme arising in the Hubbard model for contributions to the functional integral that contain quantum derivatives. Such integrals are always present in the general functional integral for the Hubbard model.
To calculate (\ref{Qseries}), in the general form, we expand it in the Taylor series in the derivative Dq and expand the exponential $[x]_{q}$ in the series \cite{Katz}:

\bigskip \begin{equation} \int \limits_{0}^{c}f(x)e^{b[x]_{q}}dx=\sum \limits_{m=0}^{\infty }\sum
\limits_{n=0}^{\infty }\frac{D_{q}^{m}(c)f(c)}{[m]_{q}!}\frac{b^{n}}{n!}\int
\limits_{0}^{c}(x-c)_{q}^{m}[x]_{q}^{n}dx \label{QQseries} \end{equation}

\bigskip Here $(x-c)_{q}$\bigskip $^{m}=(x-c)(x-qc)(x-q^{2}c).....(x-q^{m-1}c)$, and the quantum factorial is

$$[n]_{q}!=[n]_{q}[n-1]_{q}....[2]_{q}[1]_{q}[0]_{q}$$
The constant $c$, which determines the limit scaling in the problem, arises in (\ref{QQseries}). We represent the action of an arbitrary power of $D_{q}$ on $f(x)$ as a series,

$$D_{q}^{2n}(x)f(x)=\sum \limits_{k=-n}^{n}d_{2k}^{2n}(q)f(q^{2k}x)$$
A similar formula describes the action of an odd power of the quantum derivative on a function; the summation then ranges odd powers. Integral (\ref{QQseries}) can then be expressed as a series in $f(cq^{m})$  of the form (we present only the first terms of the series)
\bigskip $f(c)[I_{0}+\frac{d_{0}^{2}}{[2]_{q}!}I_{2}+\frac{d_{0}^{4}}{%
[4]_{q}!}I_{4}+\frac{d_{0}^{6}}{[6]_{q}!}I_{6}+...]+f(cq)[\frac{d_{1}^{1}}{%
[1]_{q}!}I_{1}+\frac{d_{1}^{3}}{[3]_{q}!}I_{3}+\frac{d_{1}^{5}}{[5]_{q}!}%
I_{5}+...]+f(cq^{-1})[\frac{d_{-1}^{1}}{[1]_{q}!}I_{1}+\frac{d_{-1}^{3}}{%
[3]_{q}!}I_{3}+\frac{d_{-1}^{5}}{[5]_{q}!}I_{5}+...]+$

$\bigskip $

$f(cq^{2})[\frac{d_{2}^{2}}{[2]_{q}!}I_{2}+\frac{d_{2}^{4}}{[4]_{q}!}I_{4}+%
\frac{d_{2}^{6}}{[6]_{q}!}I_{6}+...]+f(cq^{-2})[\frac{d_{-2}^{2}}{[2]_{q}!}%
I_{2}+\frac{d_{-2}^{4}}{[4]_{q}!}I_{4}+\frac{d_{-2}^{6}}{[6]_{q}!}I_{6}+...]$

\bigskip

Here, the quantum number \ \bigskip $\lbrack x]_{q}=\frac{q^{x}-q^{-x}}{q-q^{-1}}$ \ tend to $x$ as $q->1$. We have the representation

$$I_{n}=\int \limits_{0}^{c}(x-c)_{q}^{m}\exp (b[x]_{q})dx$$
for the quantities $I_{n}$. These integrals reduce to integrals of the form $\int \limits_{0}^{c}x^{m}\exp
(b[x]_{q})dx$  and can be evaluated as infinite series in $b$ (the detailed scheme for calculating integrals of this type and detailed expressions for the series in these integrals will be published elsewhere). As a result, we have $I_{n}\approx 1$ \ as \ $q->0.$ We present expressions for several quantum numbers:
 $\lbrack 1]_{q}=\frac{q^{1}-q^{-1}}{q-q^{-1}}=1$;

$[2]_{q}=\frac{q^{2}-q^{-2}}{q-q^{-1}}=q+q^{-1}$;

$[3]_{q}=\frac{q^{3}-q^{-3}}{q-q^{-1}}=q^{2}+1+q^{-2}$
We note that (\ref{QQseries}) reduces to (\ref{series})  as $q\rightarrow 1$. Expression (\ref{QQseries}) therefore contains a standard perturbative scheme for determining a functional integral. We consider the opposite limit $q<<1$. The expressions for quantum numbers and their factorials in this limit are \ $\ [1]_{q}\sim 1 $ , $\ [2]_{q}\sim q^{-1}$, \
$[3]_{q}\sim q^{-2}$ , \ $[4]_{q}\sim q^{-3}$ , $\ [5]_{q}\sim q^{-4}$ , \ $[n]_{q}\sim q^{-(n-1)}$  for the general term.

Expressions for the factorials for $q<<1$ \ , are: \ \bigskip $\lbrack 1]_{q}!=1$ ,
\ $\ [2]_{q}!\sim q^{-1}$ , \ \ \bigskip $\lbrack 3]_{q}!\sim q^{-3}$ , \ $\ [4]_{q}!\sim q^{-6}$ ,
\ $\ [5]_{q}!\sim q^{-10}$ , \ and: \ $[n]_{q}!\sim
\frac{1}{q^{n-1}}\frac{1}{q^{n-2}}...\frac{1}{q^{2}}\frac{1}{q}$ for the general term.
Polynomials in the small-$q$ regime are :$ p_1^1=-q;\ p_2^2=q;\ p_3^3=1;\ p_4^4=1/q^2;\ p_5^5=-1/q^5;\ p_6^6=1/q^6 .$ 
Substituting these expressions for $q<<1$  in formula (\ref{QQseries}), we obtain the series

$$\int_0^c f(x)e^{b[x]_{q}}dx \sim f(c)+f(cq)q+ f(cq^{2})q^2 + f(cq^{3})q^3 +f(cq^{4})q^4
+f(cq^{5})q^5+ .. $$
We find that expression (\ref{QQseries}) becomes the Jackson integral \cite{Katz} in the leading approximation in $q$. It is known \cite{Vladimirov}, that for $q=\frac{1}{p}$ , where $p$ is a prime, the Jackson integral becomes the p-adic integral.

\section{ Conclusion}
We have shown the existence of the regime described by the functional integral with the Jackson measure and deformation parameter $q$ in the Hubbard model. At $q=1/p$ , in accordance with  \cite{Vladimirov}, this regime is described by the p-adic integral. To demonstrate the features of this approach, we consider the

\begin{figure}[ht]%
\centering
\includegraphics*[width= 200pt, height=121pt]{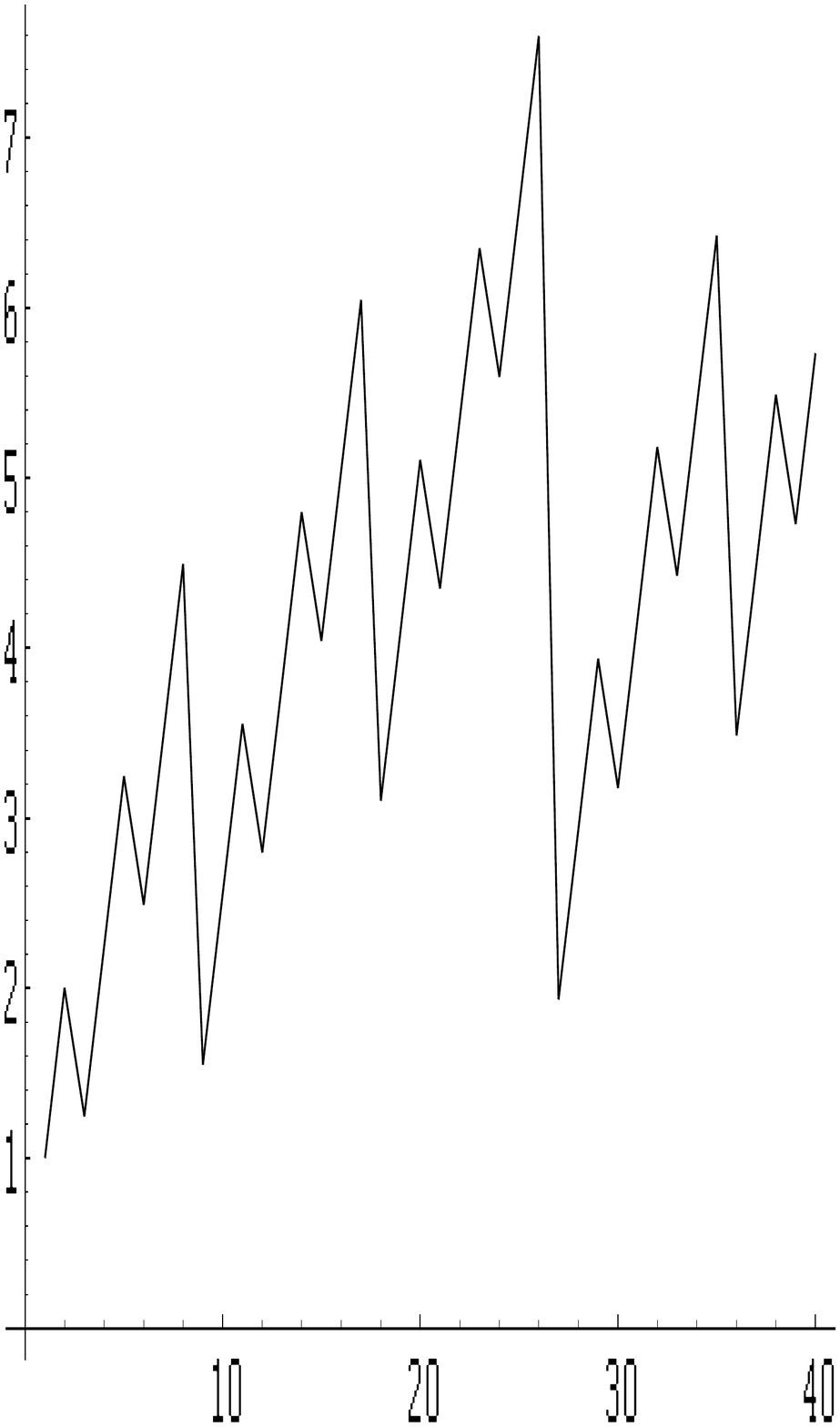}
\caption{Fig. 1.    A subcritical wave for b < 1.
}
\label{onecolumnfigure}
\end{figure}

following problem. As proposed in \cite{cc1,cc2}, we describe the conductivity by the function $ f_b(r)={\sum}_{0}^{N }$ $a_{k}p^{bk};\quad
a_{k}=(0,1,....,p-1);k\in Z.$  
 The coefficients  $ a_k$ are here equal to the coefficients in the p-adic number representation defined by the series r$r={\sum}^{N }_{0}a_{k}p^{k};\quad
a_{k}=(0,1,....,p-1);k\in Z.$
We briefly describe the necessary information concerning p-adic numbers: $p$ is a prime; an arbitrary rational number $r$ admits the representation $r=p^{\nu }\frac{m}{n}$ where $n$ and $m$ are coprime to $p$. The p-adic norm of a rational number is  $\mid r\mid _{p}=p^{-\nu }$, $\mid 0\mid _{0}=0$. The field of the p-adic numbers $ Q_p$  is the completion of the field of rational numbers $Q$ with respect to the p-adic norm.
We take $p = 3$. The numerical modeling of this function for $b = 0.5$ and $b = 1.5$ is shown in the figures. On the ordinate, we have integers presented as series for $r$. The coefficients in $ f_b(r)$  are taken from the p-adic expansion for these numbers. The result for $ f_b(r)$  is presented on the axis OY. We set $b = 0.5$ in Fig. 1 and $b = 1.5$ in Fig. 2.
In Fig. 3, we present the dependence of the conductivity on the voltage for the TA-PPE polymer. It is clear that the proposed function of the p-adic argument qualitatively describes two types of steps in this compound.

We have shown that the regime described by the p-adic functional integral exists in the functional integral for the Hubbard model.

\begin{figure}[ht]
\centering
\includegraphics*[width=200pt, height=121pt]{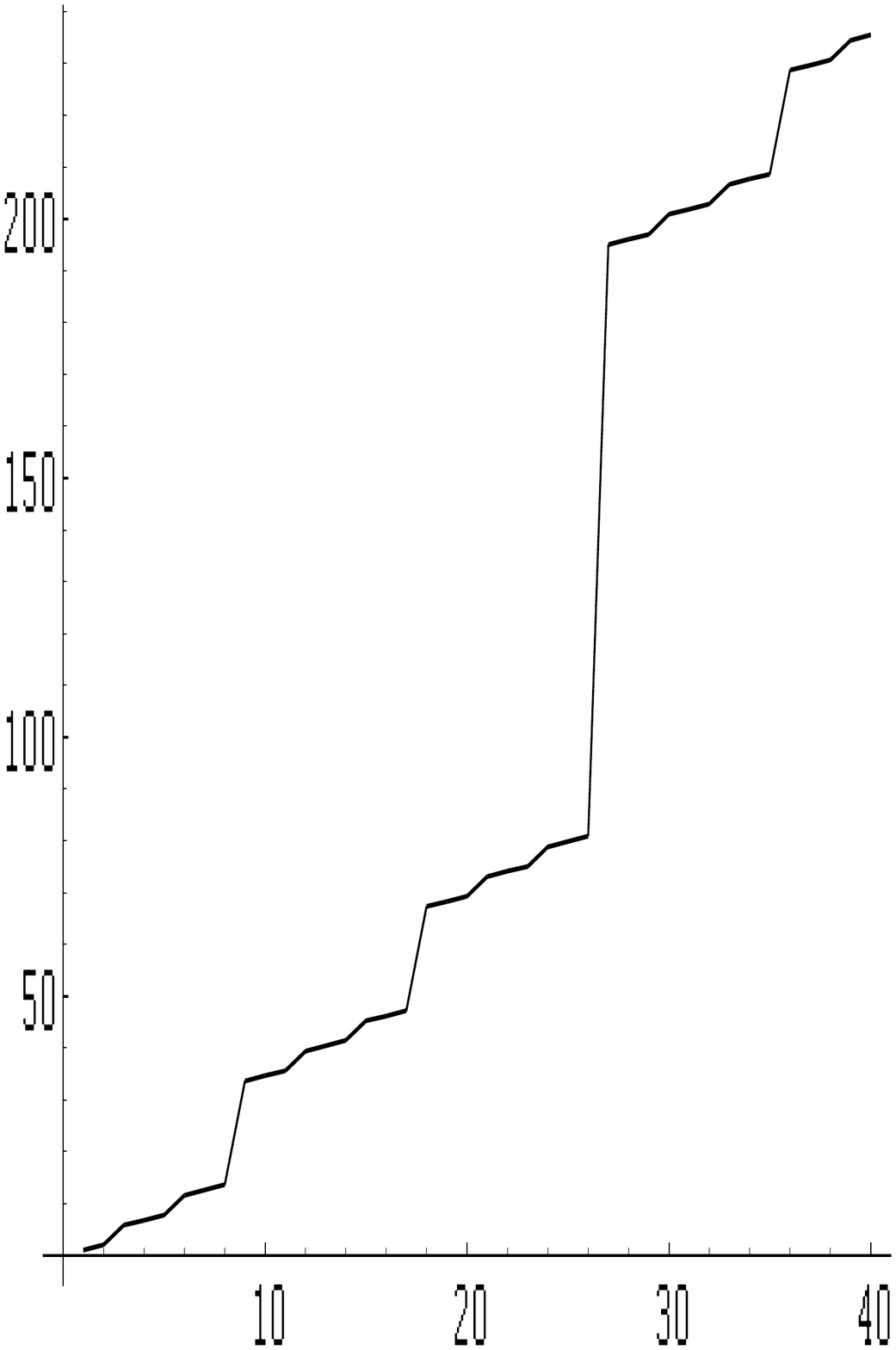}
\caption{Fig. 2.    A supercritical wave for b > 1.
}\label{fig:name5}
\end{figure}

Finally, we stress that the functional integral in the Hubbard model, as shown in this paper, can be successfully reformulated in terms of both the standard and the Jackson integral and that the latter can be reduced to the p-adic integral in a particular case. We have the impression (supported by the comparison with experimental data \cite{c11,c12,c13,c14}) that in the strong correlation regime, the description of multi-electron systems is invariant under the choice of the number field. We can begin with either a p-adic or real number description. It seems plausible that this situation can be clarified if an exact analytic calculation of coefficients in expressions (\ref{QQseries}) would be possible. Because our main regimes were \ $q->1$\ and \ $q->0$, it seems plausible that we should study the crossover regime from the p-adic to the standard perturbation series based on Gaussian integration in more detail if we will be able to calculate series (\ref{QQseries}) more exactly. This regime would clarify an interesting question: which order parameter appears in the system upon the introduction of the p-adic description whose distinct feature is the appearance of fractal behavior in physical quantities?

Acknowledgments. The author thanks the referee whose questions helped to improve the paper substantially.

\end{document}